\documentclass[review]{elsarticle}

\usepackage{lineno}
\usepackage{caption}
\usepackage{subcaption}
\usepackage{amsmath}
\usepackage{graphicx}

\usepackage[%
    colorlinks=true,
    linkcolor=red
]{hyperref}

\journal{Journal of \LaTeX\ Templates}

\begin{document}

\begin{frontmatter}

\title{Waveform Simulation for Scintillation Characteristics of NaI(Tl) Crystal}

\author[a]{J.J.~Choi}
\author[b]{C.~Ha}
\author[c]{E.J.~Jeon}
\author[c]{K.W.~Kim}\ead{kwkim@ibs.re.kr}
\author[a]{S.K.~Kim}
\author[c,d]{Y.D.~Kim}
\author[c]{Y.J.~Ko}\ead{yjko@ibs.re.kr}
\author[b]{B.C.~Koh}
\author[c,d]{H.S.~Lee}
\author[d,c]{S.H.~Lee}
\author[a]{S.M.~Lee}
\author[d,c]{B.J.~Park}
\author[e,c]{G.H~Yu}

\address[a]{Department of Physics and Astronomy, Seoul National University,\\Seoul 08826, Republic of Korea}
\address[b]{Department of Physics, Chung-Ang University,\\ Seoul 06973, Republic of Korea}
\address[c]{Center for Underground Physics, Institute for Basic Science (IBS),\\Daejeon 34126, Republic of Korea}
\address[d]{IBS School, University of Science and Technology (UST), \\Daejeon 34113, Republic of Korea}
\address[e]{Department of Physics, Sungkyunkwan University, Suwon 16419, Republic of Korea}

\begin{abstract}
The lowering of the energy threshold in the NaI detector is crucial not only for comprehensive validation of DAMA/LIBRA but also for exploring new possibilities in the search for low-mass dark matter and observing coherent elastic scattering between neutrino and nucleus. Alongside hardware enhancements, extensive efforts have focused on refining event selection to discern noise, achieved through parameter development and the application of machine learning. Acquiring pure, unbiased datasets is crucial in this endeavor, for which a waveform simulation was developed. The simulation data were compared with the experimental data using several pulse shape discrimination parameters to test its performance in describing the experimental data. Additionally, we present the outcomes of multi-variable machine learning trained with simulation data as a scintillation signal sample. The distributions of outcomes for experimental and simulation data show a good agreement. As an application of the waveform simulation, we validate the trigger efficiency alongside estimations derived from the minimally biased measurement data.
\end{abstract}


\end{frontmatter}

\section{Introduction}
Since DAMA/LIBRA claimed the observation of an annual modulation signal attributed to dark matter (DM)~\cite{BERNABEI1998195,Bernabei:2013xsa,Bernabei:2018jrt}, no signal supporting their observation has been substantiated, and the DAMA/LIBRA signal has been disfavored by many experiments based on various models~\cite{wrro120971,PhysRevLett.118.101101,Aprile:2018dbl,COSINE-100:2021xqn,PhysRevD.98.062005}. Several research groups are presently conducting experiments using the same NaI(Tl) crystal detector as DAMA/LIBRA in attempts to replicate and test the signal~\cite{kwkim15,Adhikari:2017esn,Antonello2021,Fushimi:2021mez,anais2016,cosinus2023}. Recently, COSINE-100 and ANAIS-112 published results from their model-independent annual modulation analyses~\cite{PhysRevD.106.052005,Amare:2021yyu}. Although inconclusive at present, more definitive results are expected to be reached within the next few years.

When interpreting those results, one noticeable challenge is lowering the analysis threshold. DAMA/LIBRA reported quenching factor measurements of 0.3 for sodium and 0.09 for iodine~\cite{BERNABEI1996757}, whereas recent studies have reported lower values~\cite{Collar:2013gu, PhysRevC.92.015807,Joo:2018hom,Bignell_2021,cintas2024measurement,lee2024measurements}. Discrepancies may stem from differences in measurement methodologies, although the possibility of crystal-dependence cannot be entirely dismissed. Considering such discrepancies, a direct comparison that encompasses the entire range of modulation observed by DAMA/LIBRA might necessitate a threshold twice as low as theirs. While a comprehensive comparison of the entire modulation range is not imperative for validation or refutation of the modulation signal, a sufficiently low analysis threshold is advantageous for reducing the required exposure duration~\cite{Zurowski2022}. Additionally, DAMA/LIBRA have recently presented the annual modulation results with an energy threshold down to 0.75\,keV~\cite{DAMA075}.

The low energy threshold presents additional opportunities, such as the search for weakly interacting massive particles (WIMPs) and the observation of coherent elastic neutrino-nucleus scattering (CE$\nu$NS). As sodium and iodine are both odd-proton elements, a synergy with the low energy threshold provides competitive sensitivity for spin-dependent interactions of low-mass WIMPs with protons~\cite{COSINE-100:2021poy}. Simultaneously, there are efforts to observe CE$\nu$NS of reactor neutrinos using NaI detectors~\cite{neon2023}. This experiment aims to enhance our understanding of reactor neutrinos and provide complementary insights into CE$\nu$NS involving relatively lightweight nuclide. The essential element of the CE$\nu$NS observation lies in achieving an extremely low energy threshold. The efforts toward lowering the energy threshold from a hardware perspective have brought significant enhancements in encapsulation design~\cite{Choi:2020qcj}, and concurrent endeavors on the software front via low-energy event analysis need to be pursued alongside.

To improve the analysis of low-energy events, we devised a waveform simulation reflecting the scintillation characteristics and incorporating a data acquisition (DAQ) system. This simulation contributes significantly to refining event selection aimed at lowering the energy threshold by enhancing our understanding of detector responses. This is demonstrated through the development of parameters for pulse shape discrimination (PSD) and their utility in generating pure samples for multi-variable machine learning (ML) applications.

Additionally, the analysis of extremely low-energy events must account for the fraction of triggered scintillation events, referred to as trigger efficiency.
Measuring trigger efficiencies requires obtaining minimally biased samples, independent of triggers, and as free from noise as possible. We utilized samples generated from multiple $\gamma$\,rays emitted by a $^{22}$Na radioactive source for this purpose.

The waveform simulation was optimized to accurately replicate the experimental data and is detailed in Sec.~\ref{sec:sim}, along with its implementation. In Sec.~\ref{sec:psd}, we perform a comparative analysis between the experimental and simulation data using several PSD parameters to validate the simulation. Furthermore, we present the outcomes of utilizing the simulated data as a training sample for ML. The dataset used in Secs.~\ref{sec:sim} and \ref{sec:psd} was taken from the NEON experiment, with comprehensive details of the experimental setup available in Ref.~\cite{neon2023}. Meanwhile, another experimental setup for measuring trigger efficiencies is outlined in Sec.~\ref{sec:trgeff}, accompanied by an explanation of the estimation method. We also present trigger efficiencies derived from noise-free simulated data compared to real data for enhanced confidence.

\section{Waveform Simulation (WFSim)}
\label{sec:sim}
We have developed a simulation tool named {\tt{WFSim}} to thoroughly comprehend the detector response and model pure samples of scintillation events. Typically, a full simulation for a scintillation calorimeter involves emulating the detector, including the optical process of scintillation photons, and the DAQ process. However, due to the resource-intensive nature of tuning the simulation and generating events, our focus was on crafting a fast and uncomplicated simulation tool tailored specifically to our requirements.

The detector simulation utilizing {\sc Geant}4~\cite{geant4} yields the energy deposited by the interaction of particles. Photons are then generated according to the light yield of the scintillator, which is the number of photons produced relative to the energy deposited. The generated photons go through an optical process and hit the PMT's photocathode, which creates photoelectrons~(PEs) based on its quantum efficiency. Here, the number of photoelectrons generated within an event is referred to as the {\it{NPE}}, and the number of photoelectrons generated relative to the energy deposited is defined as the ``effective light yield".

{\tt{WFSim}} focuses on photoelectrons instead of photons, bypassing the optical process. To do this, we derived the expectation value of {\it{NPE}} from experimental data, and a value of {\it{NPE}} serves as input to the {\tt{WFSim}} for each event, drawn randomly from a Poisson distribution with the expectation value as the mean. Subsequently, {\tt{WFSim}} generates waveforms with the same number of data points and time interval as the experimental data.

The process of computing physics quantities, such as charge and PSD parameters, remains consistent because the simulated waveforms share the same number of data points and the time intervals between them as the experimental data.
This allows direct comparison between the physics quantities derived from the simulation data and those from the experimental data, enabling the simulation data to function as scintillation signal samples for ML or deep learning (DL). By employing {\tt{WFSim}}, we can efficiently and precisely simulate the detector response, significantly supporting research and facilitating the development of effective analysis methods for NaI(Tl)-based experiments.

\subsection{PMT Simulation}
\label{sec:PMTsim}
The photomultiplier tube (PMT) serves to amplify PEs emitted upon photon interaction with the photocathode. This amplification, occurring through multiple dynode stages, generates a large quantity of electrons from the PEs, eventually converted into electrical pulses upon reaching the anode.
Typically, a single PE~(SPE) is multiplied by a factor of $10^6$ to $10^7$, referred to as the PMT gain $G$.
Various factors, such as voltage differences, material properties, and dynode geometry, influence the amplification factor at each dynode stage. While these factors may vary across stages in real-world scenarios, {\tt{WFSim}} assumes these as constant values for simplicity.

To stochastically determine the number of electrons amplified at stage $i$ for an electron from stage $i-1$, it is assumed to follow a Poisson distribution,
\begin{equation}
    f_\mathrm{amp}(n_{e,i}|\,G,\,n_\mathrm{stage}) = f_\mathrm{pois}(n_{e,i}|\,A_e = G^{1/n_\mathrm{stage}}) = \frac{A_e^{n_{e,i}}}{n_{e,i}!}e^{-A_e}.
\label{eq:PMTpois}
\end{equation}
where $n_{e,i}$ denotes the number of electrons amplified at the $i^\mathrm{th}$ stage, $n_\mathrm{stage}$ is the number of dynode stages, and $A_e$ represents the amplification factor of each stage. We assume each stage has the same amplification factor. For each electron obtained in stage $i-1$, the count of electrons amplified in stage $i$ is randomly drawn from Eq.~\ref{eq:PMTpois}, and their sum constitutes the total electron count in stage $i$,
\begin{equation}
    N_{e,i} = \sum^{N_{e,i-1}}n_{e,i}^\mathrm{gen},
\label{eq:nesum}
\end{equation}
where $n_{e,i}^\mathrm{gen}$ represents the random number generated via Eq.~\ref{eq:PMTpois} for each electron in stage $i-1$.

SPE pulses are often under-amplified due to cathode/dynode stage skipping or inelastic backscattering~\cite{SALDANHA201735}. Incorporating under-amplified SPE pulses termed as low-gain (LG) pulses, {\tt{WFSim}} assumes that these LG pulses have skipped one dynode without amplification. We employ the same $A_e$ as normal-gain (NG) pulses but reducing the number of stages to $n_\mathrm{stage}-1$. Each SPE pulse randomly chooses either NG and LG based on the measured probability of LG for each PMT, and the PMT amplification process is executed iteratively following Eq.~\ref{eq:nesum}.

\begin{figure}[tb]
\begin{center}
\includegraphics[width=0.9\textwidth]{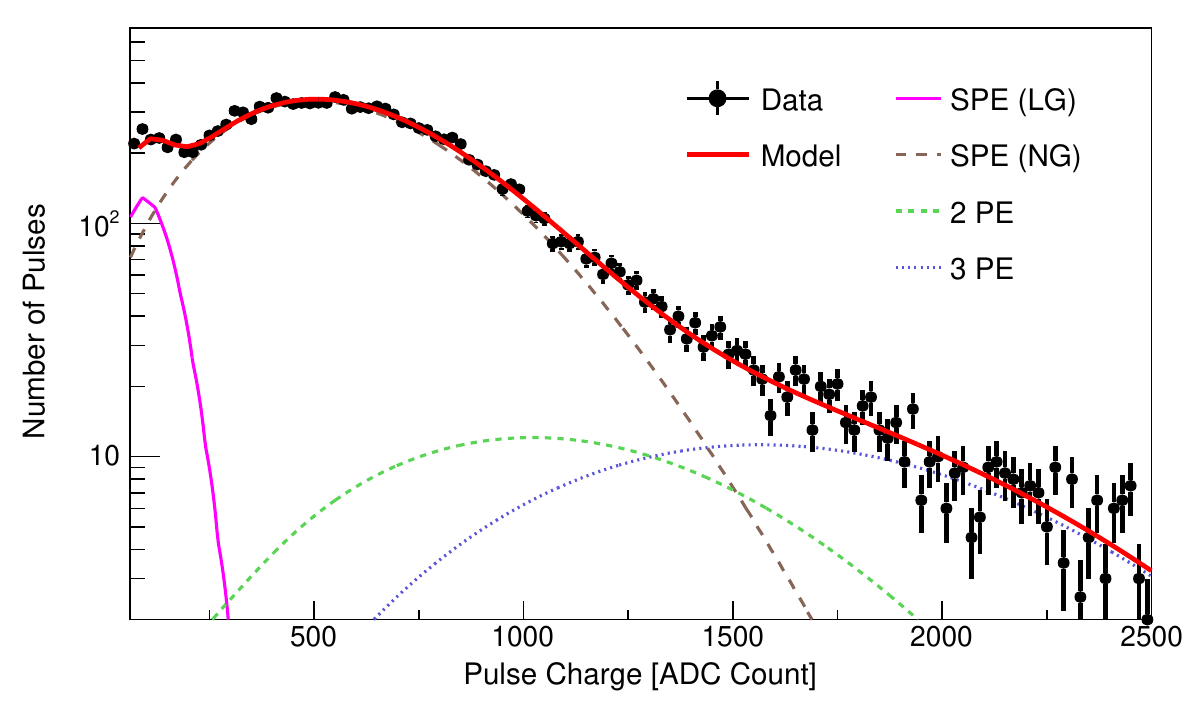} 
\caption{Pulse charge distribution with the {\it{N}}-PE model. The distribution of experimental data (black dots) of 59.5-keV $\gamma$\,rays from a $^{241}$Am source is represented by black dots with error bars, while the solid thick red line illustrates the {\it{N}}-PE model as described in Eqs.~\ref{eq:xn} to \ref{eq:y}. The solid magenta, long dashed brown, dashed green, and dotted blue lines represent distributions for low-gain (LG), normal-gain (NG), 2-PE, and 3-PE pulses, respectively. The 2-PE and 3-PE distributions encompass both NG and LG pulses, maintaining an identical fraction of LG pulses.}
\label{fig:lowgain}
\end{center}
\end{figure}

LG fraction integrated into the simulation is estimated from experimental data through modeling. The dataset used for the analysis is obtained by exposing the NaI(Tl) crystal to a $^{241}$Am $\gamma$\,ray source, and PE pulses are identified from the waveform via a clustering algorithm introduced in Ref.~\cite{LEE2006201}. To select isolated PE pulses, 59.5-keV $\gamma$\,ray events were selected and a criterion was set with a time window ranging from 2.6\,$\mu$s after to 4.6\,$\mu$s after the trigger time. The charge distribution of those pulses can be depicted in Fig.~\ref{fig:lowgain}, where the ADC count is a charge unit and is covered in detail in Sec.~\ref{sec:dig_trg}. The model to describe the data should reflect the characteristics of the NG and LG pulses. Notably, pulses extracted from the data may not exclusively comprise SPE pulses. Overlapping of multi-PE (MPE) pulses may occur making them indistinguishable and potentially biasing the results, so that we should account for this effect.

\begin{figure}[tb]
\begin{center}
\includegraphics[width=0.9\textwidth]{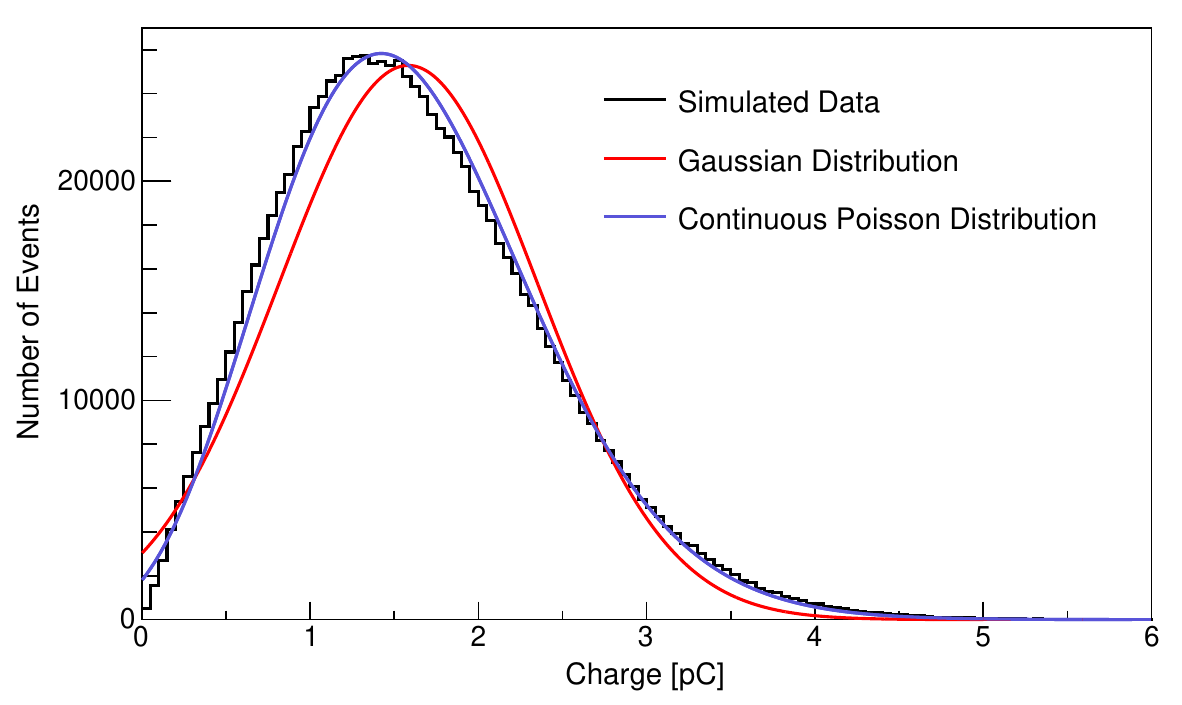} 
\caption{SPE charge distribution from PMT simulation. The red and blue lines are the Gaussian and continuous Poisson distributions modeled on the simulation results, respectively.}
\label{fig:spefit}
\end{center}
\end{figure}

The Gaussian and continuous Poisson distributions were considered for modeling the pulse charge distribution. Figure~\ref{fig:spefit} shows the two model candidates and the charge distribution from the PMT simulation based on Eq.~\ref{eq:nesum}. The continuous Poisson distribution using the Euler Gamma function was chosen because it better describes the shape of the simulated distribution and was used to model the charge distribution of the single-NG pulse,
\begin{equation}
    f_{X_N}(q|\,a_q,\,\mu_n) = f_\mathrm{CP}(q/a_q|\,\mu_n)=\frac{\mu_n^{(q/a_q)}}{\Gamma(q/a_q+1)}e^{-\mu_n},
\label{eq:xn}
\end{equation}
where $q$ represents the charge sum of the waveform and $a_q$ is a scaling factor. A random variable $X_N$ formalizes $q$ to describe the charge distribution of NG pulses with $\mu_n$ depicting the Poisson mean of $q/a_q$. With the ratio of the mean charge of NG to LG pulses $\rho_{NL}$, the charge distribution of single-LG pulses is described as,
\begin{equation}
    f_{X_L}(q|\,a_q,\,\mu_n,\,\rho_{NL}) = f_\mathrm{CP}(q/a_q\cdot \rho_{NL}|\,\mu_n).
\end{equation}
This expression assumes similarity in the charge distribution shapes for LG and NG pulses, as their shapes are primarily determined by the first amplification stage while subsequent stages smear the distribution.

The model to describe MPE pulses should also account for the characteristics of NG and LG pulses. For MPE pulses composed of solely NG or LG pulses, a Poisson distribution with $\mu_n$ multiplied by the number of PEs as the mean can be used. When MPE pulses are a mix of NG and LG pulses, a distribution with the combination of $X_N$ and $X_L$ as the random variable is needed for modeling. For instance, a distribution like the following can be applied for a 2-PE charge distribution consisting of an NG pulse and an LG pulse,
\begin{equation}
    f_Y(q) = \int f_{X_N}(t|\,a_q,\,\mu_n)\times f_{X_L}(q-t|\,a_q,\,\mu_n,\,\rho_{NL})\,dt,
\label{eq:y}
\end{equation}
where $Y$ is the random variable representing the sum of $X_N$ and $X_L$~($Y = X_N + X_L$).

By extending this approach to encompass the charge distribution up to 3-PE pulses, the model adequately represents the data as shown in Fig.~\ref{fig:lowgain}. The estimated fraction of LG pulses from this model fit is ($6.79\pm0.46$)\%, and the effective light yield given the fraction can be estimated as,
\begin{equation}
    NPE/\mathrm{keV} = \frac{Q(59.5\,\mathrm{keV})/(59.5\,\mathrm{keV})}
        {(1-f_\mathrm{LG})\,q_\mathrm{NG} + f_\mathrm{LG}\,q_\mathrm{LG}},
\end{equation}
where $Q(59.5\,\mathrm{keV})$ is the charge corresponding to the 59.5-keV $\gamma$\,ray from $^{241}$Am and $f_\mathrm{LG}$ denotes the LG fraction. The mean charge of NG and LG pulses are represented by $q_\mathrm{NG}$ and $q_\mathrm{LG}$, respectively. These can also be obtained from the model fit as $q_\mathrm{NG} = \mu_na_q$ and $q_\mathrm{LG} = \mu_na_q/\rho_{NL}$. The information such as SPE gain, LG pulse probability, and effective light yield is used as input to the simulation. These processes are conducted individually to reflect the characteristics of each PMT and the results are utilized in the PMT simulation.

\subsection {Waveform Generation}
Waveform generation in {\tt{WFSim}} involves the placing SPE and random pulses on the pedestal followed by the analog-to-digital converter's (ADC) sampling. The following subsections explain the implementation of these components and the extraction of essential data-derived information.

\subsubsection {Pedestal}
The waveform is discretized with a time interval of 2\,ns, aligning with the 500-MHz ADC sampling rate of our DAQ system. Each time bin's pedestal value is drawn randomly from a Gaussian distribution,
\begin{equation}
    f_\mathrm{ped}(q|\,\mu_\mathrm{ped},\,\sigma_\mathrm{ped}) = \frac{1}{\sqrt{2\pi}\sigma_\mathrm{ped}}\exp\left[{-\frac{(q-\mu_\mathrm{ped})^2}{2\sigma_\mathrm{ped}^2}}\right],
\end{equation}
where $\mu_\mathrm{ped}$ and $\sigma_\mathrm{ped}$ denote the mean and root-mean-square (RMS) of the pedestal, respectively. During data processing for each event, the pedestal mean is computed and subtracted from the waveform before deriving physics quantities. Therefore, maintaining a constant $\mu_\mathrm{ped}$ does not introduce any difference from the experimental data.

\begin{figure}[bt]
\begin{center}
\includegraphics[width=0.9\textwidth]{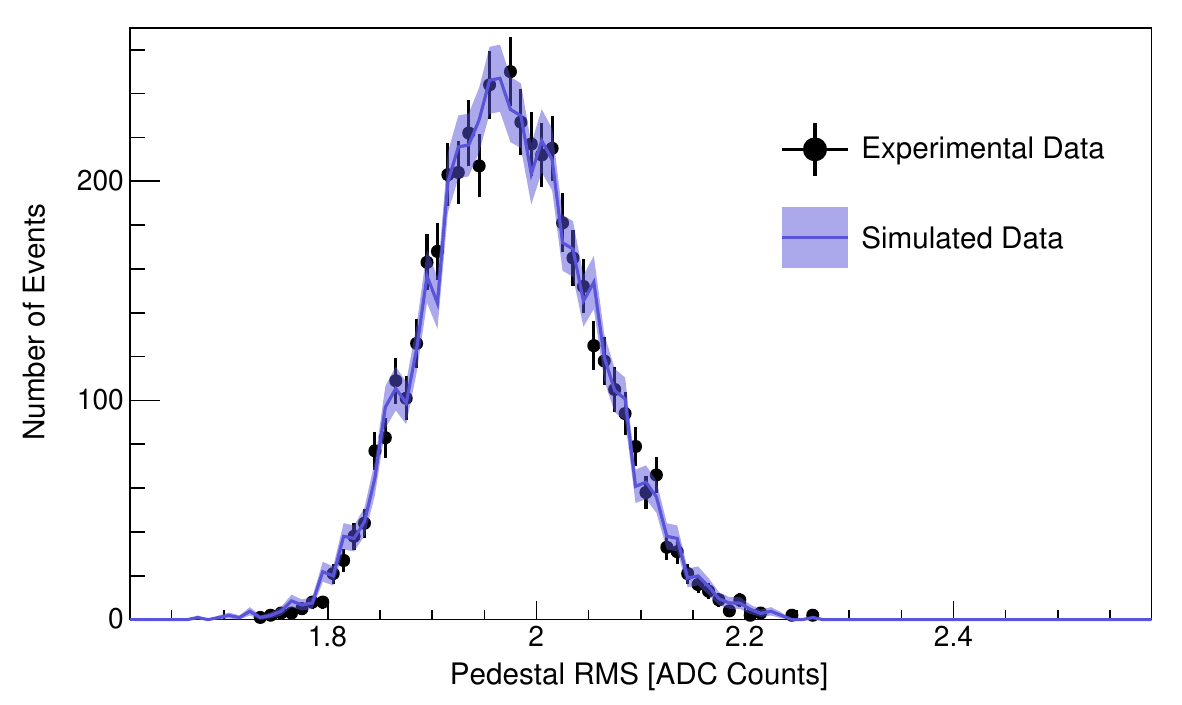} 
\caption{Comparison of the pedestal RMS distributions between the experimental data (black dots) and the simulated data (solid blue line), which agree very well.}
\label{fig:pedrms}
\end{center}
\end{figure}

Conversely, if the pedestal RMS differs from the experimental data, it may impact the waveform; thus, it should mirror the data. Given the asymmetric nature of the experimental data's pedestal RMS distribution, we employed an asymmetric Gaussian distribution to model the variations of $\sigma_\mathrm{ped}$,
\begin{equation}
    f_\mathrm{\sigma_\mathrm{ped}}(\sigma_\mathrm{ped}|\,\mu_\sigma,\,\sigma_l,\,\sigma_r) =
    \begin{cases}
        \frac{2}
        {\sqrt{2\pi}\left(\sigma_l+\sigma_r\right)}\exp{\left[-\frac{(\sigma_\mathrm{ped} - \mu_\sigma)^2}{2\sigma_l^2}\right]}&(\sigma_\mathrm{ped} < \mu_\sigma),\\
        \frac{2}
        {\sqrt{2\pi}\left(\sigma_l+\sigma_r\right)}\exp{\left[-\frac{(\sigma_\mathrm{ped} - \mu_\sigma)^2}{2\sigma_r^2}\right]}&(\sigma_\mathrm{ped} \geq \mu_\sigma),
    \end{cases}
\label{eq:pedrms}
\end{equation}
where $\mu_\sigma$ denote the most probable value of $\sigma_\mathrm{ped}$. The dispersion of $\sigma_\mathrm{ped}$ is represented by $\sigma_l$ and $\sigma_r$ when it is less or greater than $\mu_\sigma$, respectively. For each event, $\sigma_\mathrm{ped}$ is randomly drawn from the distribution in Eq.~\ref{eq:pedrms} to generate the pedestal values. These equation parameters were deduced from the experimental data to feed into the simulation. Figure~\ref{fig:pedrms} illustrates a comparison between the simulation and experimental data for the pedestal RMS distribution, showing a good agreement between the experimental and simulation data.

\subsubsection {SPE Pulse Generation}
\label{sec:spepulse}
The initial step in generating an SPE pulse involves selecting the pulse type, either NG or LG pulses, followed by obtaining the SPE charge. By default, the SPE charge is acquired through the PMT simulation outlined in Sec.~\ref{sec:PMTsim} for each event. Alternatively, templates obtained from 100 million PMT simulations, including both NG and LG pulses, can be utilized to avoid repeated PMT simulations. In this approach, the SPE charge is randomly extracted from the template corresponding to each pulse type.

\begin{figure}[tb]
\begin{center}
\includegraphics[width=1.0\textwidth]{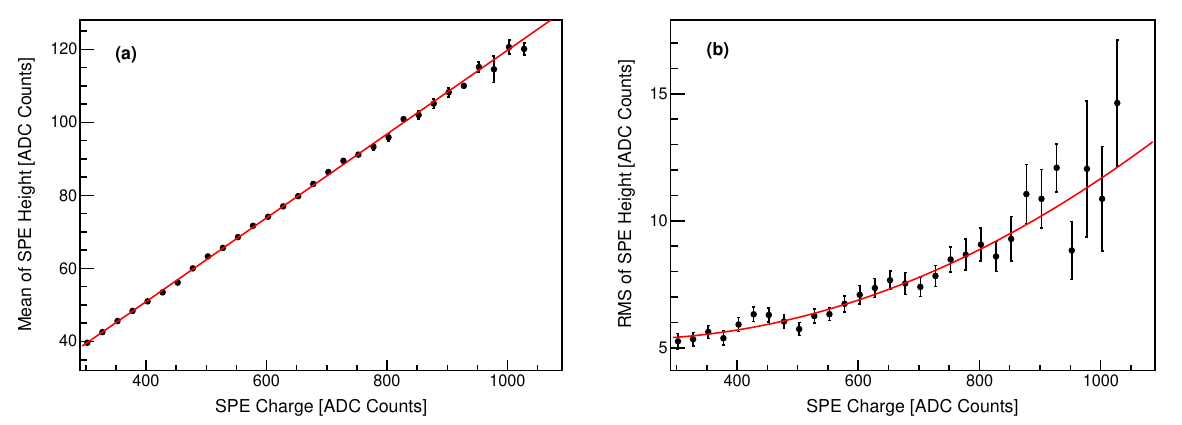} 
\caption{(a) Mean and (b) RMS of the SPE height distribution as functions of the SPE charge. The black dots with error bars depict parameters estimated using a Gaussian model fitted to experimental data. These parameter values are modeled with first and second-order polynomials (red lines) representing the variations in the mean and RMS as the SPE charge changes, respectively.}
\label{fig:spe_qh}
\end{center}
\end{figure}

The SPE pulse's height, the maximum amplitude of the pulse, is then determined using a charge-height relationship.

Since the SPE height distribution for a given SPE charge was observed to be Gaussian-like shape in the experimental data, the distribution is modeled by a Gaussian distribution varying its mean and standard deviation depending on the SPE charge $q$,
\begin{equation}
    f_h(h|\,q) = f_\mathrm{gaus}[h|\,\mu_h(q),\,\sigma_h(q)],
\label{eq:height_gaus}
\end{equation}
where $h$ represents the SPE height and $\mu_h$ stands for the mean of the SPE height distribution corresponding to a given $q$, while $\sigma_h$ represents the standard deviation. The values of $\mu_h$ and $\sigma_h$ in the Gaussian model vary with the change in SPE charge indicating variations in the distribution as demonstrated in Fig.~\ref{fig:spe_qh}. Those values are respectively modeled with first and second-order polynomial functions $\mu_h(q)$ and $\sigma_h(q)$, representing the charge-height relationship. Upon randomly generating the SPE charge, the height distribution is determined and the SPE height is drawn from this relationship.

\begin{figure}[tb]
\begin{center}
\includegraphics[width=0.9\textwidth]{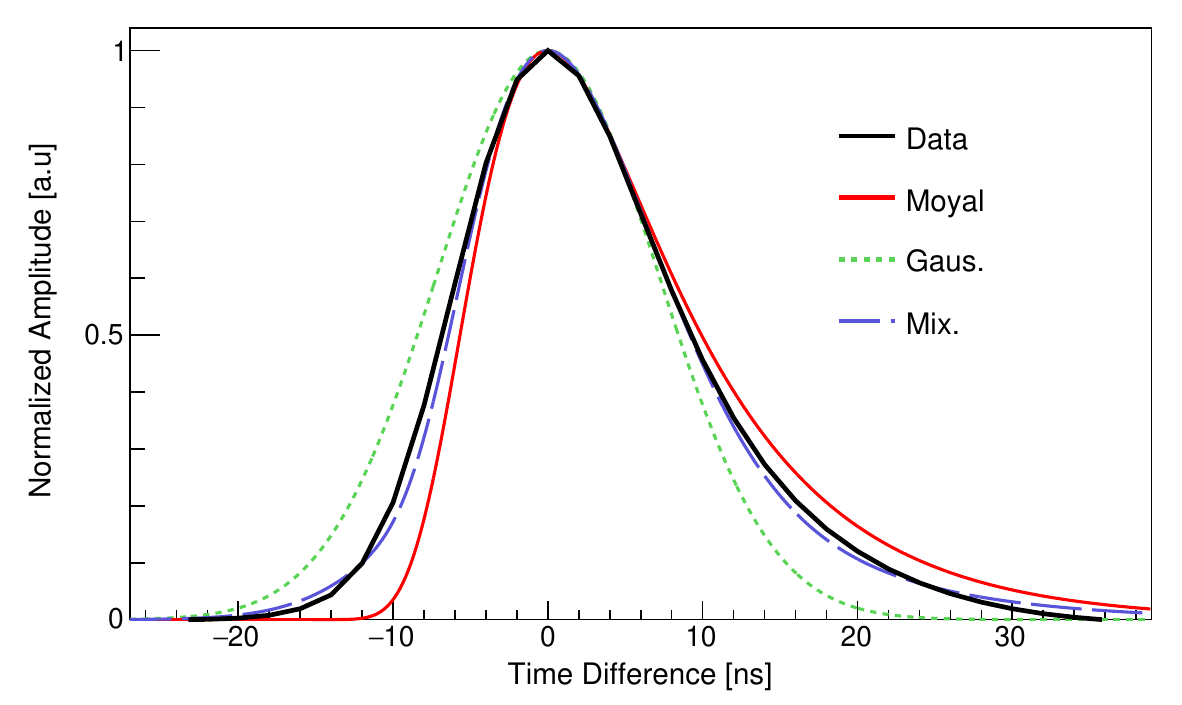} 
\caption{SPE shape distributions. We accumulated isolated pulses zeroing out the time when they maximize for data distribution. The black line denotes experimental data while the solid red, dotted green, and dashed blue lines represent the Moyal, Gaussian, and mixture models, respectively. The heights of all distributions were normalized to 1, and the widths of the reference distributions were area scaled to be equal to the data.}
\label{fig:spe_shape}
\end{center}
\end{figure}

To establish the SPE shape for a given SPE charge and height, several reference distributions were compared to the data. Isolated SPE pulses were accumulated to get the data distribution. To synchronize the pulse times before accumulating, we set the time of maximum pulse amplitude to zero. The reference models used are Gaussian, Moyal~\cite{doi:10.1080/14786440308521076}, and their mixture model expressed as,
\begin{equation}
\begin{split}
        &f_\mathrm{mix}(t|\,t_\mathrm{spe},\,\sigma_\mathrm{spe})\\ &=
    \frac{1}{\sqrt{8\pi}\sigma_\mathrm{spe}}
    \left\{
        \exp{\left[-\frac{1}{2}\left(\frac{t - t_\mathrm{spe}}{\sigma_\mathrm{spe}} + e^{-\frac{t-t_\mathrm{spe}}{\sigma_\mathrm{spe}}}\right)\right]}
        +\exp\left[-\frac{(t-t_\mathrm{spe})^2}{2\sigma_\mathrm{spe}^2}\right]
    \right\},
\end{split}
\label{eq:speshape}
\end{equation}
where $t_\mathrm{spe}$ denotes the SPE hit time, and $\sigma_\mathrm{spe}$ represents the dispersion of the SPE pulse. As shown in Fig.~\ref{fig:spe_shape}, the mixture model was chosen because it explains the shape such as asymmetry better than the two other models. Once the charge and height of the SPE are determined, the SPE pulse is generated at the specified SPE hit time detailed in the subsequent subsection.

\subsubsection {SPE Hit Time}
Obtaining the hit time of each SPE typically involves characteristics of scintillation process such as rising and decaying times. However accurate simulation of this process necessitates a comprehensive understanding of the detector response, including scintillation process and optical properties of the NaI(Tl) crystal. It also demands significant simulation time. To bypass these without missing the scintillation characteristics of NaI(Tl) crystal, {\tt{WFSim}} introduced an alternative method by extracting hit times from a data-driven time distribution generated through a two-step process.

\begin{figure}[tb]
\begin{center}
\includegraphics[width=0.9\textwidth]{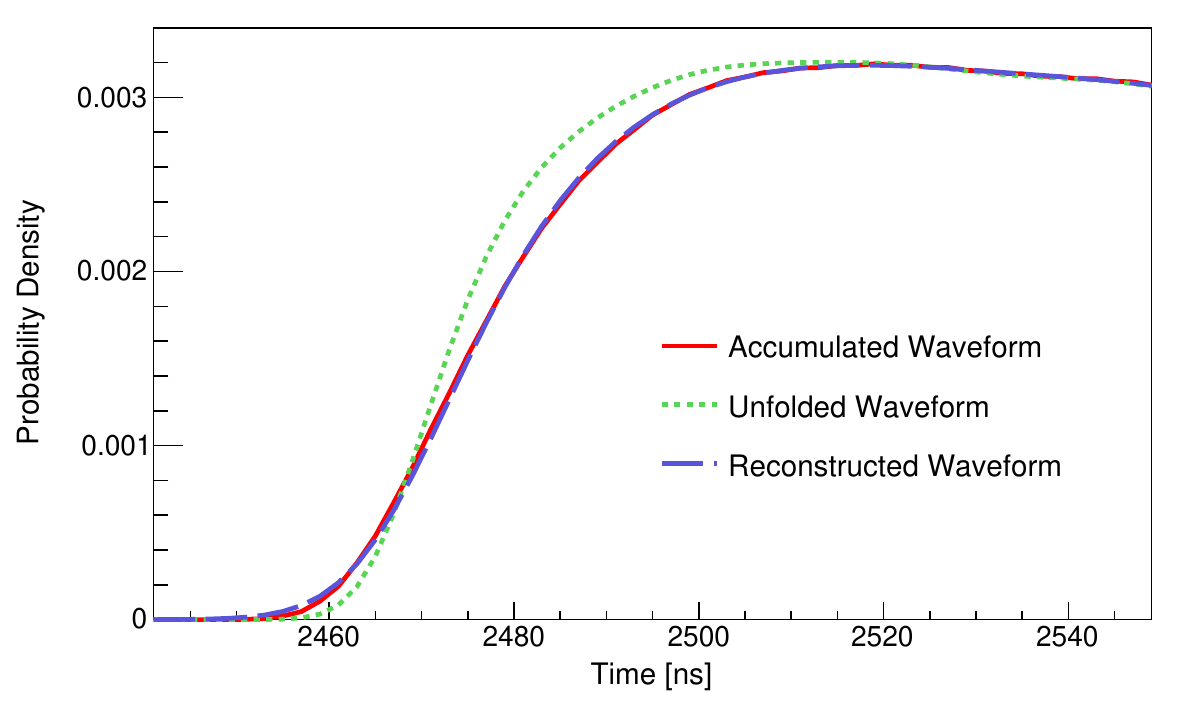} 
\caption{Waveform distribution comparison. The solid red line illustrates the waveform accumulated from scintillation events in experimental data, the dotted green line represents the waveform unfolded from the accumulated waveform, and the dashed blue line shows the waveform reconstructed from the unfolded waveform.}
\label{fig:unfold}
\end{center}
\end{figure}

First, an average waveform was derived by accumulating scintillation events within the [2,\,10]\,keV energy range, selected due to its representation of a relatively low-energy area where noise is easily discriminated. To remove the smearing effect caused by the time distribution of the SPE pulses, the iterative Bayesian unfolding (IBU) method~\cite{DAGOSTINI1995487} was employed. The IBU method unfolded the accumulated waveform through a migration matrix, which can be obtained from simulations outlined in Sec.~\ref{sec:spepulse} as follows,
\begin{enumerate}
    \item Generate hit time randomly within [2000,~2002]\,ns to account for the 2\,ns time interval of the waveform.
    \item Input the generated hit time into the simulation to get the waveform.
    \item Accumulate the waveforms obtained by repeating the simulation 100,000 times and normalize the accumulated waveform to obtain a PDF, which is the probability distribution of the output over the input of [2000,~2002]\,ns.
    \item The PDF for [t, t+2] ns is obtained by shifting the PDF of 3.
\end{enumerate}

The unfolded waveform was reconstructed via the migration matrix to be validated,
\begin{equation}
    w_i^\mathrm{recon} = \sum_jM_{ij}\,w_j^\mathrm{unfold}
\end{equation}
where $w_i^\mathrm{recon}$ denotes the amplitude of reconstructed waveform for $i^\mathrm{th}$ time bin while $w_j^\mathrm{unfold}$ represents the amplitude of unfolded waveform for $j^\mathrm{th}$ time bin, and the migration matrix is indicated by $M_{ij}$. Figure~\ref{fig:unfold} shows that the reconstructed waveform (dashed blue) closely matches the accumulated waveform (solid red). The unfolded waveform is also depicted in the figure as a dotted green line. The $t_{spe}$ values in Eq.~\ref{eq:speshape} are randomly drawn from the unfolded waveform to determine the position of each pulse.

\subsubsection {Random Photoelectrons}
{\tt{WFSim}} includes random PEs originating from sources such as NaI(Tl) phosphorescences~\cite{PhysRevD.107.032010,PhysRevD.93.042001} and PMT dark current. Given that the waveform recording in the experimental data starts roughly 2.4\,$\mu$s before the trigger time similar as the COSINE-100 DAQ system~\cite{Adhikari_2018}, the rate of random PEs can be estimated by counting pulses within the [0,\,2]\,$\mu$s timeframe. This does not include cases that are accidentally triggered by random PEs, but since they are on the order of 0.12\% for the estimated 6000-Hz rate, they are negligible. These random PEs are assumed to share the same shape and magnitude as the SPE pulses. The position is randomly chosen to ensure an uniform distribution.

\subsection {Digitization and Triggering}
\label{sec:dig_trg}
The PMT generates an analog signal measured in volts, which is then digitized by the ADC. The DAQ system used in the COSINE-100~\cite{Adhikari_2018} and the NEON~\cite{neon2023} experiments has a resolution ($\sigma_\mathrm{ADC}$) of 12\,bits and a peak-to-peak voltage ($V_\mathrm{pp}$) of 2.5\,V. The amplitude of the signal is therefore quantized at a voltage resolution of $2.5/2^{12}$\,V ($\sim0.61$\,mV). This quantization occurs at a 500-MHz sampling rate, discretizing the signal at 2-ns intervals. Upon satisfying specific local trigger conditions with the digitized signal, the ADC transmits a trigger decision to the trigger control board (TCB). Given that a NaI(Tl) crystal integrates two PMTs to read scintillation light, the TCB generates a global trigger decision when both PMTs' trigger decisions meet a time coincidence criterion. Upon generating the global trigger decision, the DAQ server records 8-$\mu$s digitized waveforms starting 2.4\,$\mu$s before the trigger time.

\begin{figure}[tb]
\begin{center}
\includegraphics[width=1.0\textwidth]{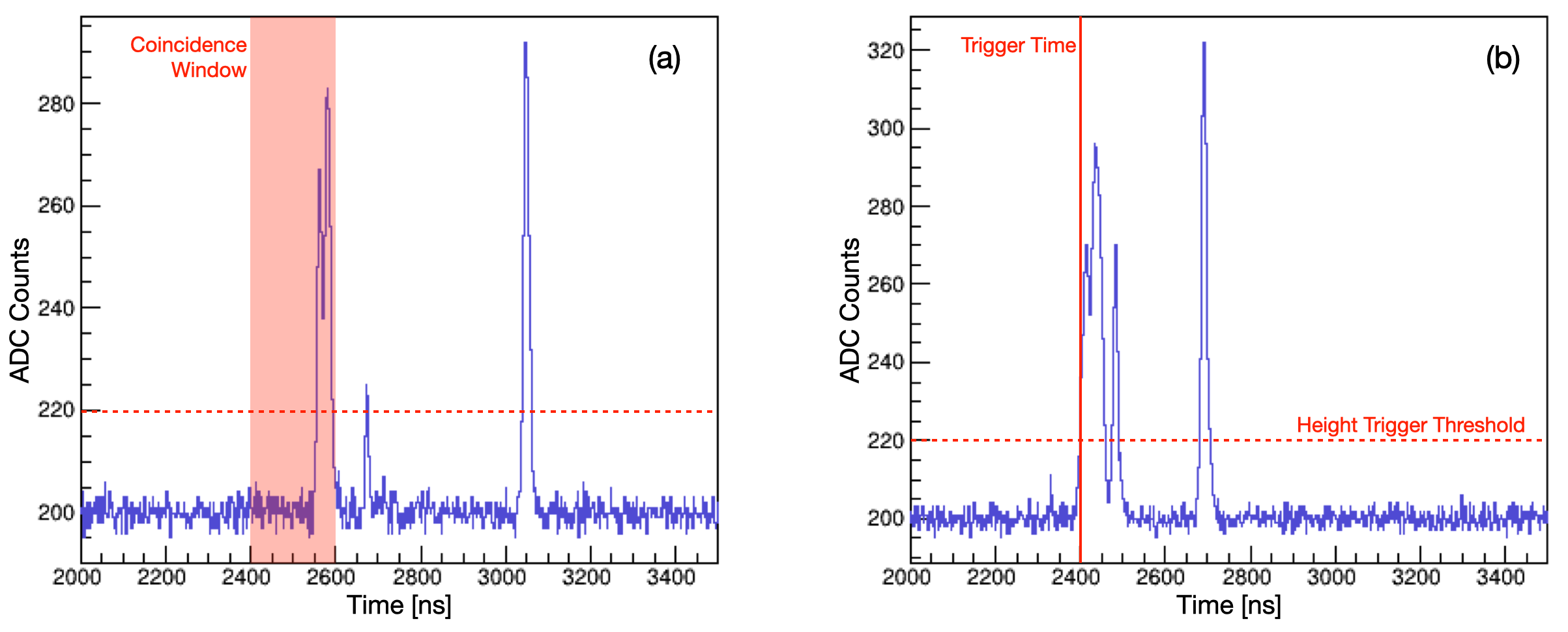} 
\caption{Waveforms generated by {\tt{WFSim}} for an event. (a) and (b) represent the waveforms for PMT channel 1 and 2, respectively. The solid red line denotes the trigger time, the dotted red line shows the height trigger threshold, and the red shaded area illustrates the coincidence window. Note that to increase the readability, the waveforms show zoomed-in view from 2000\,ns to 3500\,ns out of 8000\,ns total readout window.}
\label{fig:trigger}
\end{center}
\end{figure}

{\tt{WFSim}} includes the discretization process by generating waveforms with 2-ns time interval. The waveform then undergoes quantization and triggering processes similar to the DAQ system~\cite{Adhikari_2018}. Since the output from the PMT simulation represents the amplified number of electrons, the amplitude is initially expressed in units of charge, namely pico-Coulomb~(pC). It needs a unit conversion to match the quantization standards of the experimental data, and the amplitude is converted into ADC counts, a unit that ensures the ADC’s voltage resolution of 1. The factor $C_\mathrm{pC2ADC}$ which accounts for this conversion is expressed as,
\begin{equation}
    C_\mathrm{pC2ADC} = \frac{2^{\sigma_\mathrm{ADC}}\,R_\mathrm{ter}(\Omega)}{w_t(\mathrm{ns})\,V_\mathrm{pp}(\mathrm{V})}\times10^{-3},
\end{equation}
where $R_\mathrm{ter}$ is the terminal resistance of 50\,$\Omega$. 

\begin{figure}[tb]
\begin{center}
\includegraphics[width=1.0\textwidth]{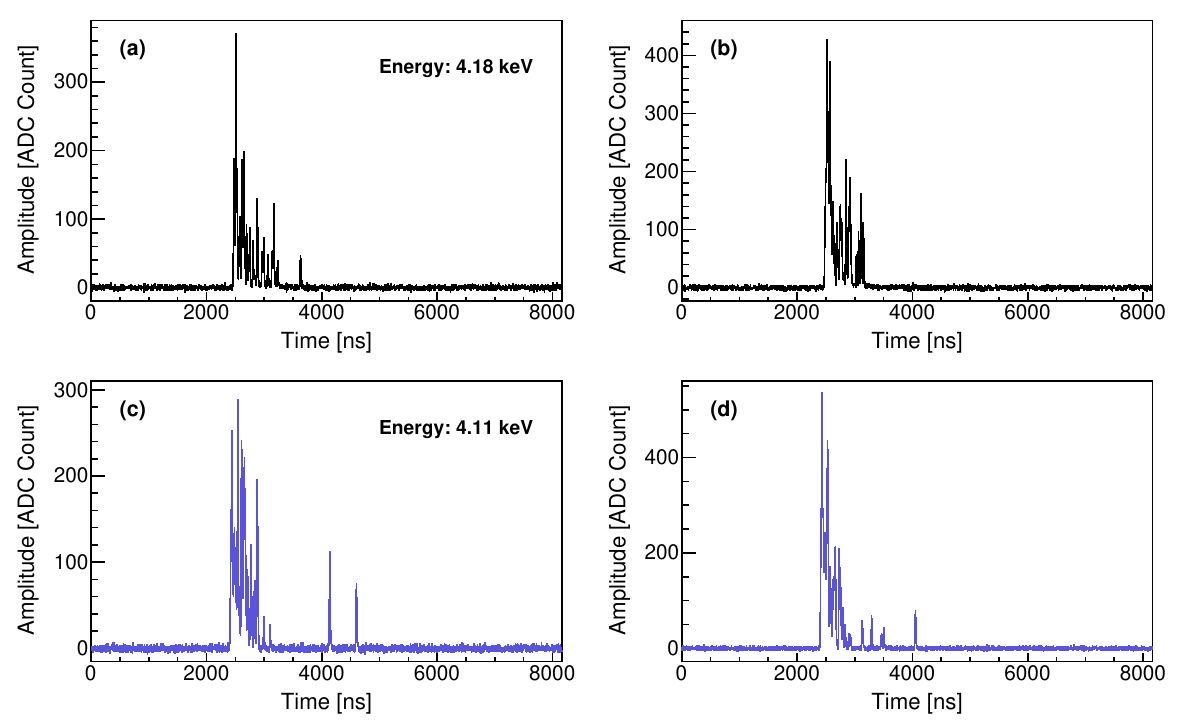} 
\caption{Waveform examples. Pedestal subtracted waveforms from experimental and simulation data are displayed in (a) (b) and (c) (d), respectively, for PMT channel 1 and 2.}
\label{fig:waveform}
\end{center}
\end{figure}

After converting the waveform amplitudes into ADC count-based values, each time bin's amplitude is quantized into an integer. Local and global triggers that are integrated into the process are implemented as follows: the local trigger deploys a height trigger condition upon amplitude threshold crossing. The global trigger opens a coincidence window (CW) when a local trigger activates on a PMT channel. If another channel triggers while the CW remains open, it flags the event as triggered. The waveform is shifted to align the trigger time with a specified time and is recorded, accompanied by a trigger bit indicating event triggering status.

Figure~\ref{fig:trigger}~(b) demonstrates the waveform on channel 2 crossing the height trigger threshold, initiating a 200-ns CW. This event is then triggered because channel 1's waveform also crosses the threshold within the CW as shown in Fig.~\ref{fig:trigger}~(a). The local trigger time on channel 2 serves as the global trigger time and is aligned to 2.4\,$\mu$s. Figure~\ref{fig:waveform} displays waveforms from experimental and simulation data as an example.

\section{Validation of WFSim}
\label{sec:psd}
The primary objective of {\tt{WFSim}} is to enhance event selection by analyzing low-energy events. Simulation plays a crucial role in developing PSD parameters by understanding the detector response to scintillation events. It is also essential in training multi-variable ML models using PSD parameters. For this purpose, the simulation must accurately represent the experimental data regarding PSD parameters, ensuring consistency in ML outputs. Here we use boosted decision tree (BDT) as used in the COSINE-100 event selection~\cite{adhikari2020lowering}. Since the simulation generates waveforms based on SPE pulses, several variables characterizing the waveform were compared between the simulation and experimental data.

\subsection{SPE Variables}
\begin{figure}[tb]
\begin{center}
\includegraphics[width=1.0\textwidth]{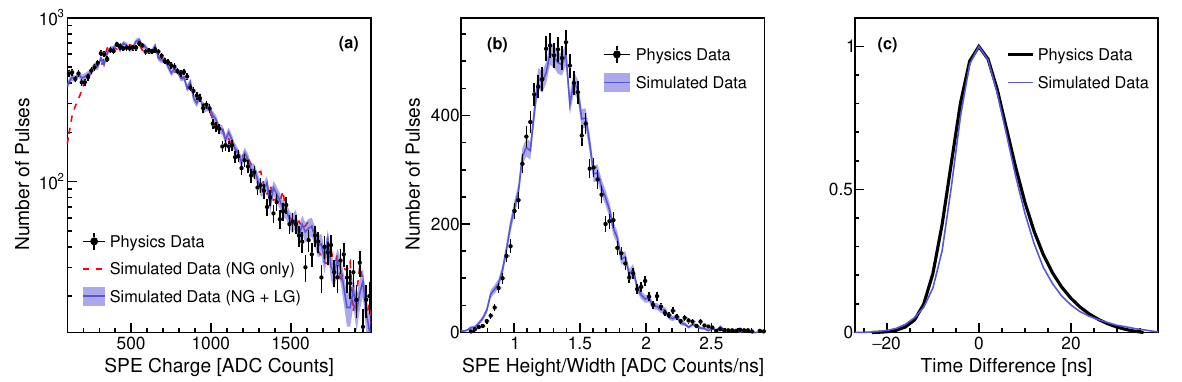} 
\caption{Three SPE distributions. (a) The pulse charge distribution. Experimental data is represented by black dots. The red line illustrates the distribution solely considering NG pulses, while the blue line represents the distribution accounting for both LG and NG pulses. (b) The pulse height-to-width distribution. Experimental data (black dots) is compared with simulation data (solid blue lines). (c) The pulse shape. Data distribution in Fig.~\ref{fig:spe_shape} compared with simulation distribution. For all distributions, the experimental data used is the same as in Fig.~\ref{fig:lowgain}.}
\label{fig:spevars}
\end{center}
\end{figure}

Figure~\ref{fig:spevars} displays three distributions of SPE variables using the dataset selected for Fig.~\ref{fig:lowgain} in Sec.~\ref{sec:PMTsim}. To generate simulation data for comparing with experimental data under the same conditions, the following process was employed: a Poisson distribution was initially defined with the {\it{NPE}} expectation value equivalent to 59.5\,keV as the mean. A randomly generated {\it{NPE}} from this distribution was then fed into {\tt{WFSim}} for each event to produce a waveform. Following identical processing steps as those applied to the experimental data, the charge distribution was obtained with isolated pulses between 2.6 and 4.6\,$\mu$s after the trigger time within their respective waveforms.

As shown in Fig.~\ref{fig:spevars}~(a), the simulation accounting for only NG pulses (red line) closely approximates the experimental data (black dots) across most regions, but exhibits a noticeable differences in the low-charge region. While we cannot be completely certain that those differences are due to LG pulses, the values obtained from the model, such as the fraction of LG pulses, are reasonable and the simulation considering both NG and LG pulses (blue line) aligns well even in the low-charge region. Obviously, if {\tt{WFSim}} is utilized in an analysis where the impact of LG pulses is important, further validation should be required. Figure~\ref{fig:spevars}~(b) demonstrates the height-to-width ratio distribution of the SPE, showing effective alignment between simulation and experimental data. Here, the SPE width refers to the time interval from the beginning to the end of the pulse. The SPE shape distribution of experimental data is also compared with that of simulation data in Fig.~\ref{fig:spevars}~(c). The simulation pulse times were synchronized in the same way as the experimental data by zeroing the time bin when the amplitude was at its maximum.

\subsection{PSD Parameters}
When a particle deposits energy in a NaI(Tl) crystal, scintillation light is emitted, generating a scintillation signal characterized by a decay time of 250\,ns~\cite{Tanabashi2018}. Conversely, PMT-induced noise (Type-I) exhibit a notably shorter decay time of 50\,ns or less. The low-energy region presents a greater challenge in discrimination, as another type of noise events (Type-II) emerges alongside typical noise events. Separating the scintillation events from these noise events is crucial to lowering the energy threshold, thus COSINE-100 has focused on event selection analysis via PSD techniques based on the waveform's characteristics. In this section, the simulation is compared to the experimental data in terms of several PSD parameters used in COSINE-100's event selection analysis~\cite{adhikari2020lowering}.

NaI(Tl) detectors in the NEON experiment were installed in liquid scintillator (LS) to reduce background by tagging radiation-related events~\cite{neon2023}. Events coinciding with LS are categorized as multiple-hit events and excluded when forming the physics-search dataset. However, as these events exhibit relatively low noise, they serve well for comparative distribution analysis with simulation. Conversely, single-hit events without LS coincidences typically contain a higher noise level, especially at lower energies, making them usable as noise samples for BDT training. The analysis in this section uses 42\,days of NEON data, with two pre-cuts: difference between the pedestal and the lowest amplitude of the waveform less than 10\,ADC counts and the second pulse time larger than 2\,$\mu$s.

\begin{figure}[tb]
    \begin{center}
        \includegraphics[width=1.0\textwidth]{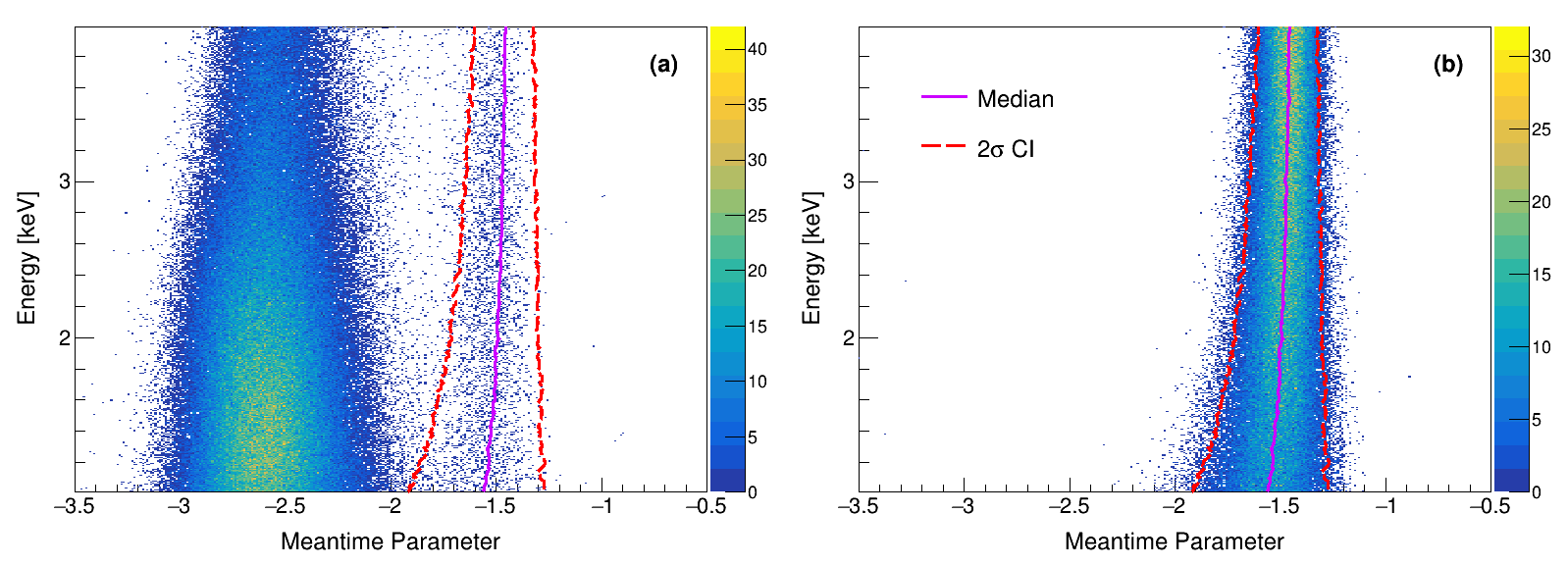}
    \end{center}
  \caption{Two-dimensional distributions between the Meantime Parameter and energy. (a) illustrates the distribution of multiple-hit events in experimental data, while (b) shows the distribution of simulation data. The solid purple and dashed red lines denote the median and 2$\sigma$ confidence interval of the simulation data.}
  \label{fig:validmpar}
\end{figure}

The charge-weighted mean time capitalizes on the differences in decay time, exhibiting strong discrimination power against Type-I noise. Figure~\ref{fig:validmpar} displays the two-dimensional distributions between the energy and the Meantime Parameter, which is a combination of charge-weighted mean times for the two PMTs as defined as
\begin{equation}
    p_m = \ln\left(\left<t\right>_1 + \left<t\right>_2\right),
\end{equation}
where $\left<t\right>_i$ is the charge-weighted mean time for PMT $i$. In Fig.~\ref{fig:validmpar}~(a) derived from experimental data, the scintillation events are distributed on the right, separated from the PMT-noise events. This distribution notably aligns with the 2$\sigma$ confidence interval from the simulation distribution, indicating the accuracy of {\tt{WFSim}} in representing the experimental data.

In addition to the Meantime Parameter, we validated the simulation by comparing them to experimental data with six PSD parameters.
\begin{enumerate}
    \item \textbf{Crystal-based Mean Time} is defined as the charge-weighted mean time of stacked waveforms from the two PMTs.
    \item \textbf{Likelihood Parameter} was key to reaching the 1-keV energy threshold in COSINE-100. This parameter is a PSD parameter based on a likelihood method, which was developed with a focus on Type-II noise discrimination~\cite{adhikari2020lowering}.
    \item \textbf{Maximum Cluster Ratio} is defined as the ratio of the charge of the largest pulse within an event to the total event charge.
    \item \textbf{Charge-to-Height Ratio} is defined as the ratio of the charge integral within the first 5\,$\mu$s of the waveform to the maximum amplitude.
    \item \textbf{Slow Charge} is defined as the ratio of the charge integrated between 100 and 600\,ns to that for the first 600\,ns.
    \item \textbf{Fast Charge} is defined as the ratio of the charge integrated over the first 50\,ns to that for the first 600\,ns.
\end{enumerate}


\begin{figure}[tb]
  \begin{center}
    \includegraphics[width=1.0\textwidth]{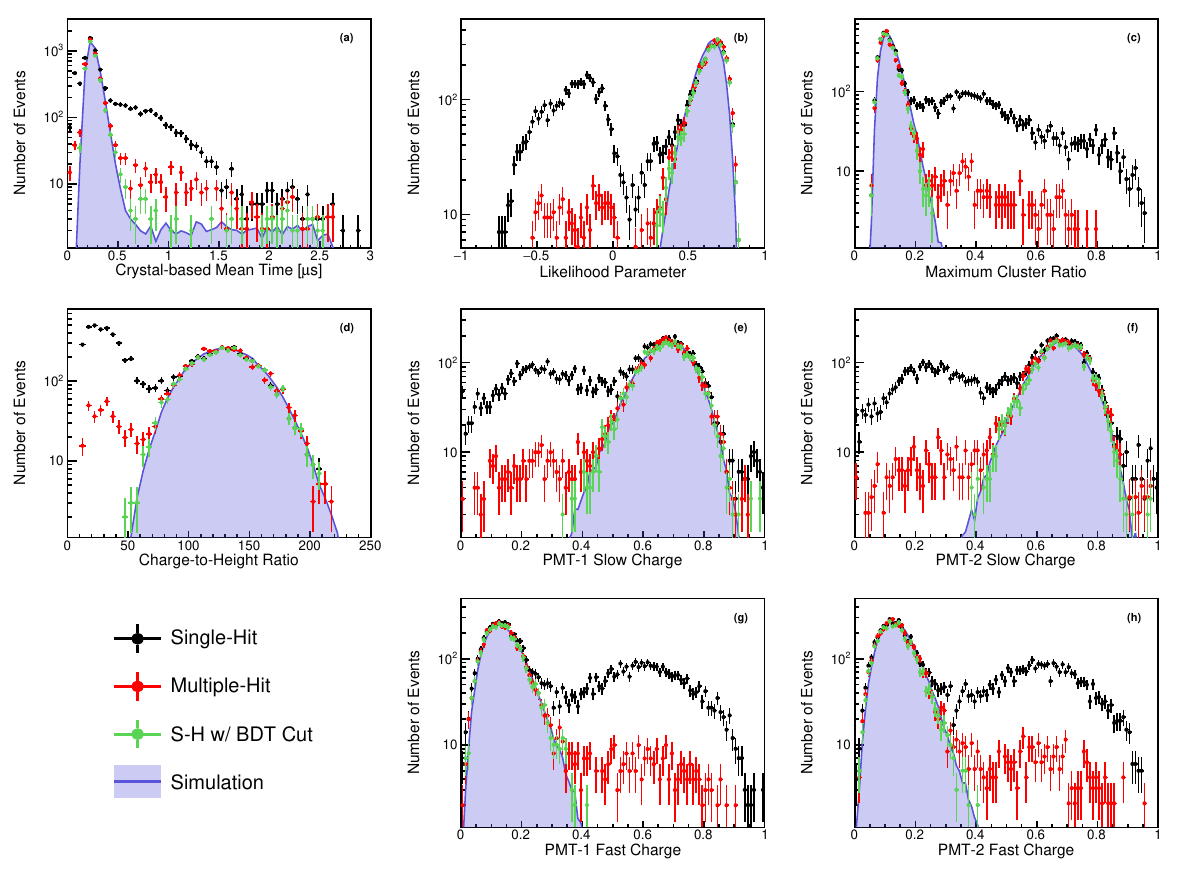}
  \end{center}
  \caption{The distributions of PSD parameters used to discriminate scintillation events from noise events. The black and red dots represent single-hit and multiple-hit events in experimental data, while blue lines denote simulation data. The green dots with error bars are single-hit distribution after applying the BDT cut. A condition was set that the Meantime Parameter must exceed -2.2 for all distributions.}
  \label{fig:validmul}
\end{figure}

Figure~\ref{fig:validmul} displays the comparisons between experimental and simulation data distributions for the aforementioned six PSD parameters. For the comparisons, experimental and simulated data were subjected to the two criteria: energies between 2 and 6\,keV and a Meantime Parameter greater than -2.2. In addition, the muon rate at the experimental site is very high, resulting in a large number of anomalous events that are presumably due to Cherenkov or scintillation effects in the PMT glass by muon. Because of this, multiple-hit events coinciding with LS events above about 35 MeV were removed. The single-hit events contain a considerable amount of noise, while notable agreement between experimental and simulation data is observed upon applying BDT conditions described later. Meanwhile, the simulation distributions align well with the multiple-hit event distributions within the scintillation signal region thanks to the relatively lower noise level. Contrary to the first four parameters, the Slow Charge and Fast Charge are PMT-based parameters, thus the distributions are shown for each PMT channel.

\begin{figure}[tb]
  \begin{center}
    \includegraphics[width=0.9\textwidth]{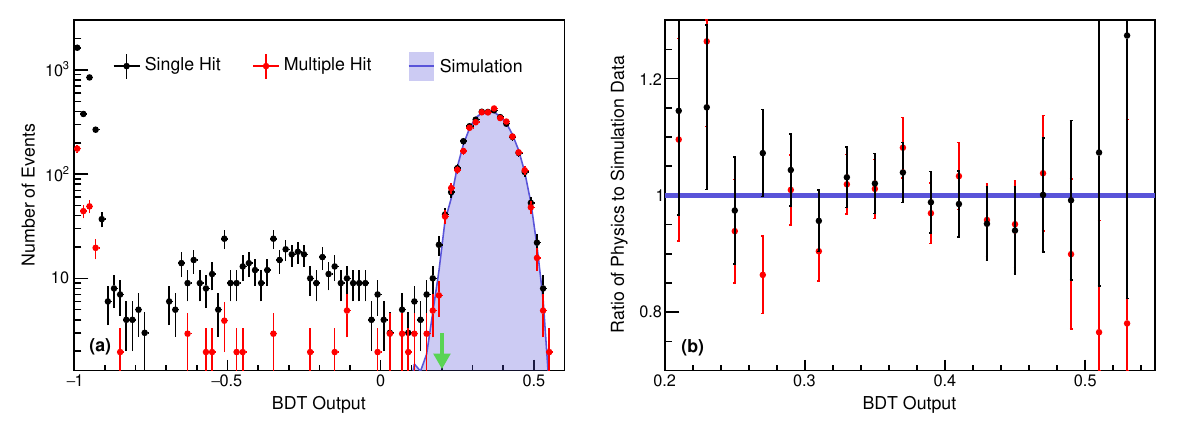}
  \end{center}
  \caption{The distribution of BDT outputs and the ratios of experimental to simulation data distribution. Black (red) dots with error bars represent single-hit (multiple-hit) events in experimental data, while the blue line denotes the simulation data. The green arrow points to the BDT cut used for Fig~\ref{fig:validmul}. A precondition was set requiring the Meantime Parameter to exceed -2.2 for all distributions.}
  \label{fig:bdtcomp}
\end{figure}

BDT stands as one of the multi-variable event selection techniques crucial in achieving a lower energy threshold by effectively distinguishing scintillation and noise events at low energy levels. In COSINE-100, BDT was utilized for event selection, which was a significant improvement~\cite{COSINE_SET1,adhikari2020lowering}. The parameters illustrated in Fig.~\ref{fig:validmul} undergo training through the BDT method and are combined into a single powerful discriminator. The simulated events were used as samples of the scintillation signal for BDT training, while the single-hit events from the experimental data were used as background (noise) samples.

As shown in Fig.~\ref{fig:bdtcomp}, the region where the BDT output is greater than 0.2 is dominated by scintillation signals and was set to the BDT conditions used in Fig.~\ref{fig:validmul}. In this region, the simulation agrees well with the experimental data and the chi-squares between them are 6.8 and 18.7 for the single-hit and multiple-hit events, respectively, with 18 degrees of freedom. The relatively large chi-square with the multiple-hit events is probably due to the fact that the events were mostly triggered by other crystals, while the events in the other data sets were all self-triggered. Consequently, {\tt{WFSim}} accurately characterizes the experimental data from various perspectives and contributes to lowering the energy threshold by favoring event selection, especially at lower energies. The potential for an unlimited number of events as samples for ML or DL purposes could lead to substantial improvements.

\section{Trigger Efficiency}
\label{sec:trgeff}
Measuring trigger efficiencies not only is essential for any meaningful physics analysis of low-energy events but also reflects the reliability of understanding the DAQ system and the detector. The main challenge in measuring efficiencies lies in sampling scintillation events that should satisfy two conditions: independence from the trigger condition and inclusion of solely scintillation events without noise events. To meet these criteria, we utilized $^{22}$Na data and selected events that met certain criteria as described later. The trigger efficiencies are then estimated from this sample of scintillation events. We also compare the efficiencies estimated via simulation with actual measurements, taking into account that {\tt{WFSim}} is implemented based on the same DAQ system used for efficiency measurements.

\subsection{Experimental Setup}
\label{sec:exp_setup}
\begin{figure}[tb]
\begin{center}
\includegraphics[width=1.0\textwidth]{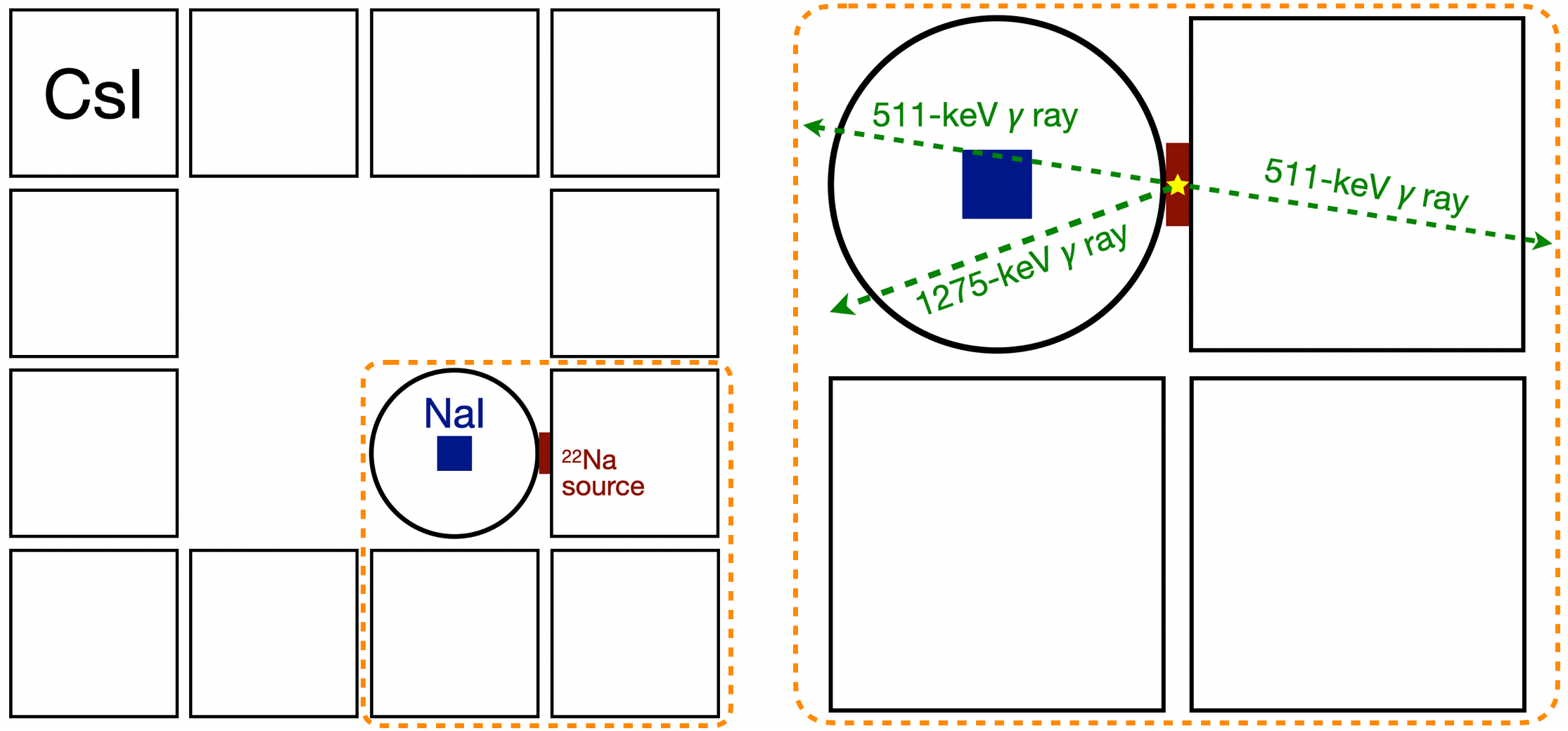}
\caption{Schematic of the experimental setup for trigger efficiency measurement. The NaI(Tl) crystal (blue filled square) was encapsulated by a copper case (black circle), and it was installed inside a twelve CsI(Tl) crystal detector array represented as black squares. $^{22}$Na radioactive source was placed between NaI(Tl) crystal and one of CsI(Tl) crystals.}
\label{fig:trgeffsetup}
\end{center}
\end{figure}

For the trigger efficiency measurement, NaI(Tl) crystals with light yield larger than 25\,PE/keV were employed to enhance trigger efficiency assessment below 1\,keV region. The crystals were assembled using a new methodology~\cite{neon2023}, directly attaching them to the PMTs, which maximized optical coupling, reduced light loss, and ensured high effective light yield. Two measurements were conducted: one on a rectangular crystal named Crystal~A (20\,mm$\times$20\,mm$\times$15\,mm) and the other on a cylindrical Crystal~B (diameter:~25\,mm,~height:~38\,mm). Crystal~A shared the same powder as Crystal~2,~5, and 8 in the COSINE-100 crystals, while Crystal~B was grown using the powder same to that used for Crystal~3 and 4 in the COSINE-100 crystals~\cite{Adhikari:2017esn}. Their effective light yields were $25.9\pm0.6$ and $27.6\pm0.3$\,PE/keV for Crystals~A and B, respectively. The measurements were conducted in the presence of the CsI(Tl) crystal array, which was used for the KIMS dark matter search experiment at the Yangyang underground laboratory~\cite{hslee07,sckim12,kwkim15} as shown in Fig.~\ref{fig:trgeffsetup}. Approximately 40\,days of data taken from April to May and from October to November in 2020 were used for the trigger efficiency estimation.

To accurately estimate the trigger efficiencies, the NaI(Tl) crystal was exposed to a $^{22}$Na radioactive source. The crystal was surrounded by twelve CsI(Tl) crystals, and the source was placed between the NaI(Tl) crystal and one of the CsI(Tl) crystals. About 90\% of $^{22}$Na decays through positron emission to the 1275-keV level of $^{22}$Ne. Thanks to the short half-life (3.7\,ps) of excited $^{22}$Ne, the $\gamma$\,ray with an energy of 1275\,keV is emitted almost simultaneously with the two 511-keV $\gamma$\,rays from electron-positron annihilation. By tagging coincident $\gamma$\,rays, where multiple $\gamma$\,rays from $^{22}$Na are detected simultaneously by NaI(Tl) and CsI(Tl) crystals, enabled the collection of minimally biased scintillation event samples for estimating the trigger efficiencies.

The NaI(Tl) and CsI(Tl) crystals each have two PMTs attached and share the trigger condition that both PMTs must be fired within the crystal specific CW. However, due to the difference in decay time of NaI(Tl) and CsI(Tl) crystals, the lengths of the CW are different, 200\,ns and 2\,$\mu$s, respectively, and there are differences in the local trigger conditions for the PMTs. The local trigger condition for NaI(Tl) crystal requires at least one PE pulse, as introduced in Sec.~\ref{sec:dig_trg}, and is consistent with the COSINE-100 and NEON experiments as well as {\tt{WFSim}}. On the other hand, the local trigger condition for CsI(Tl) crystal demands at least two PE pulses to be read by each PMT within 2\,$\mu$s. Therefore, to define an event, the global trigger is formed by a logical OR condition between the NaI(Tl) and adjacent CsI(Tl) crystals and is recorded as waveforms with a sampling rate of 500\,MHz and a length of 8\,$\mu$s.

To compare trigger efficiencies among detectors with different effective light yields, we need a variable other than energy. Since the triggering process relies on {\it{NPE}}, it is used as a proxy variable for energy. As the true {\it{NPE}} is an unknown variable, the {\it{NPE}} is converted by dividing the charge by the mean of SPE charge, obtained through modeling of the SPE charge distribution as depicted in Fig.~\ref{fig:lowgain}. This approach for the {\it{NPE}} calibration was applied equally to simulation and experimental data.

\subsection{Estimation of Trigger Efficiency}
\label{sec:est_trgeff}
As an initial step in event selection, a 511-keV $\gamma$\,ray signal was required at the CsI(Tl) crystal with timing coincidence to the NaI(Tl) crystal. Despite these conditions, there is noise contamination mainly due to direct hits of energetic $\gamma$ on PMT glasses generating Cherenkov light as well as accidental coincidences enhanced by external $\gamma$ source. Noise events are identifiable by their distinctive pulse shape characterized by significantly larger amplitudes, while randomly coincident events can cause bias as they are not identified. To mitigate the bias caused by these events, we computed a variable termed the Minimum Time Difference ($\Delta t_\mathrm{min}$). The closest time between the pulse times of the two PMTs was derived as $\Delta t_\mathrm{min}$ by identifying the pulse time of each PMT that represented the moment the waveform crossed the height trigger threshold.

\begin{figure}[tb]
\begin{center}
\includegraphics[width=0.9\textwidth]{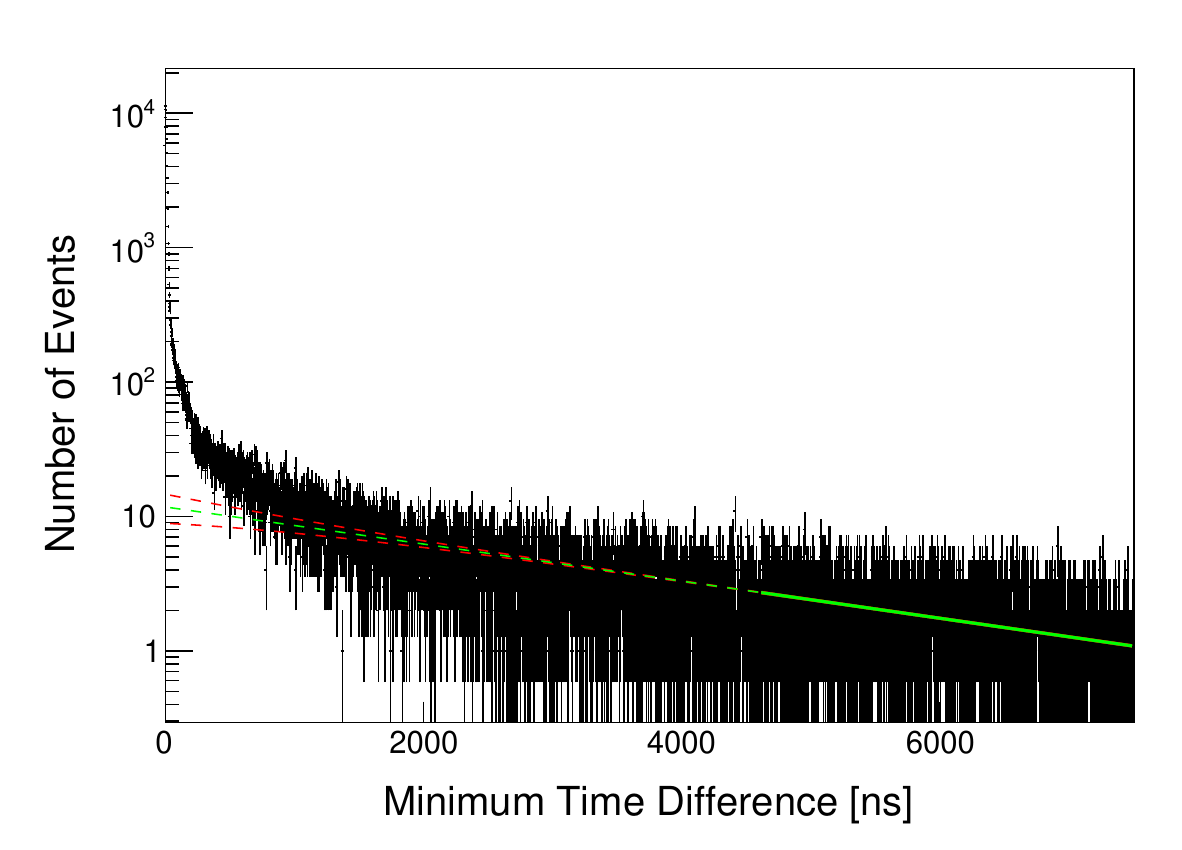} 
\caption{The distribution of the minimum time difference between the two PMTs. The green line is the fitted exponential function to data and the red lines denote its uncertainties. The solid lines show the function and its uncertainties in the fitting range, while the dashed lines represents extrapolations of those.}
\label{fig:deltat}
\end{center}
\end{figure}

As shown in Fig.~\ref{fig:deltat}, the distribution comprises short and long decay components. The short decay component indicates strong correlation between the PMTs due to scintillation events, while long decay component denotes accidental coincidence. Modeling with an exponentially decaying function in the longer decay range allowed estimation of the long decay components within the trigger region ($\Delta t_\mathrm{min}<\mathrm{CW}$) and the entire region. In Fig.~\ref{fig:deltat}, the fitted function is shown as a solid line within the range to use for modeling and a dashed line extrapolated outside the range.

There are events that have pulses in only one of the two PMTs. These are not included in Fig.~\ref{fig:deltat}, but should be included in the estimation of trigger efficiencies because they could contribute to inefficiencies. Based on the aforementioned description, trigger efficiencies can be estimated using,
\begin{equation}
    \epsilon = \frac{n_\mathrm{trg} - n_\mathrm{RC}|_{\Delta t_\mathrm{min} < \mathrm{CW}}}{(n_\mathrm{all} - n_\mathrm{RC}) + n_\mathrm{SH}},
\label{eq:trgeff}
\end{equation} 
where $n_\mathrm{all}$ denotes the number of events that has timing coincidence with the 511-keV $\gamma$\,ray signal in the CsI(Tl) crystal, while $n_\mathrm{trg}$ represents the events triggered by DAQ system. The $n_\mathrm{RC}$ stands for the estimated amount of random coincidence events and $n_\mathrm{SH}$ is the number of events registered in only one PMT.

\begin{figure}[tb]
\begin{center}
\includegraphics[width=0.9\textwidth]{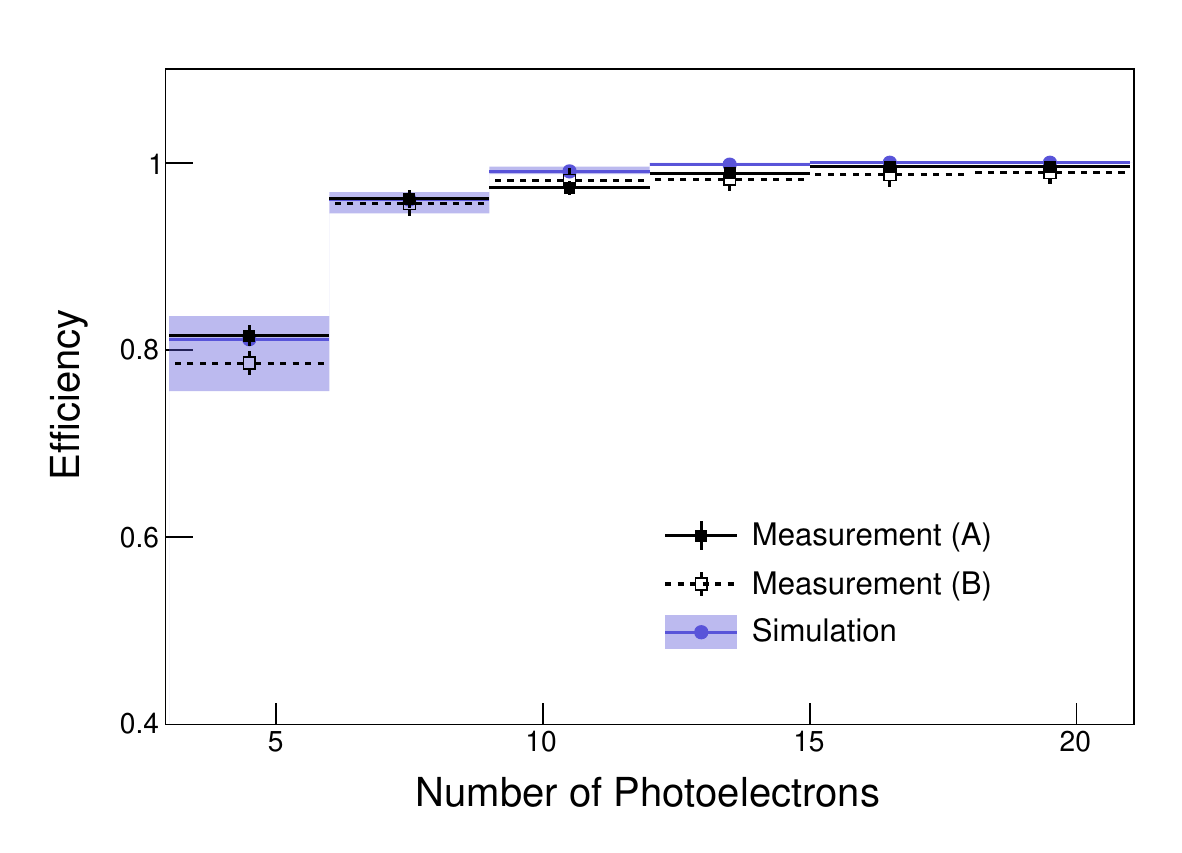} 
\caption{Trigger efficiencies measured via Crystal~A (black closed square) and B (open square) together with the simulation (blue circle) as a function of the number of photoelectrons.}
\label{fig:trgeff}
\end{center}
\end{figure}

Using {\it{NPE}} divided into six bins ranging from 3 to 21\,PEs, the trigger efficiency for each bin was estimated via Eq.~\ref{eq:trgeff}. Figure~\ref{fig:trgeff} shows the trigger efficiency measurements along with their uncertainties. The uncertainties took into account both uncertainty from noise estimation and statistical uncertainty from event counts such as $n_\mathrm{all}$ and $n_\mathrm{SH}$. The trigger efficiencies measured with Crystals A and B are consistent, revealing low dependence on crystal shapes and light emission quality. At 3 to 6\,PEs, the efficiencies are $81.5\pm1.2\%$ and $78.6\pm1.8\%$ for Crystal~A and B, respectively.

Figure~\ref{fig:trgeff} also includes a comparison with the trigger efficiencies estimated with {\tt{WFSim}} introduced in Secs.~\ref{sec:sim} and \ref{sec:psd}. With the advantage of noise-free simulation, the trigger efficiencies can be estimated simply by the ratio of triggered to generated events. To account for differences in the SPE size that could be caused by differences in the PMT's gain, the changes in trigger efficiencies for halving and doubling the amplification factor were added to the statistical uncertainties and shown as a blue filled area.

Assuming the population has a flat distribution and the probabilities of light collection by both PMTs are similar, it is expected that the trigger efficiency can be 85\% at the first bin. However, the efficiency from {\tt{WFSim}} is slightly less than this at 81.15\%, and the difference appears to be due to the CW of 200\,ns. Since the simulation was tuned to data from NEON, the dependence on the crystal might be more pronounced. Nevertheless, the agreement of the simulation with the measurements within 3\% shows a low dependence on the crystal, indicating that the results can be applied to NaI(Tl) detectors in the same DAQ environment.




\section{Summary}
The NaI(Tl) detector's low energy threshold is important for the full validation of DAMA/LIBRA, the search for low-mass DM, and observations of CE$\nu$NS. To analyze low-energy events, we have developed a data-driven waveform simulation. The simulation generates pedestals, SPE pulses, random photoelectrons, and integrates them. The PMT amplification process is simulated and the pulses are implemented using SPE variables estimated by modeling the experimental data. The waveforms from the experimental data are used to place the pulses, taking into account the scintillation characteristics of the NaI(Tl) crystal.

Our simulation was compared to the experimental data, in terms of SPE variables and PSD parameters as well as results trained with BDT technique, and showed remarkable agreement. The simulation poised to enhance future low-energy analyses in NaI(Tl)-based experiments. Plans to employ DL to train waveform for event selection underscore the simulation's critical role. To this end, we are also planning a study on the PMT afterpulse and its implementation in {\tt{WFSim}}.

Furthermore, we have measured the trigger efficiencies, which are crucial for characterizing low-energy events. A $^{22}$Na source was utilized for minimally biased samples, and the efficiency was estimated to be 80\% at 3 to 6\,PEs. Comparison with the waveform simulation presents excellent agreement proving reliability of the simulation. Our estimations from the two crystals and the simulation revealed low crystal dependency. These efforts will prove invaluable as we make efforts to achieve to an extremely low energy threshold.

\section{Acknowledgments}
We thank the Korea Hydro and Nuclear Power (KHNP) Company for providing underground laboratory space at Yangyang and the IBS Research Solution Center (RSC) for providing high-performance computing resources. This work is supported by the Institute for Basic Science (IBS) under project code IBS-R016-A1, NRF-2019R1C1C1005073, NRF-2021R1A2C3010989 and NRF-2021R1A2C1013761, Republic of Korea.

\bibliography{dm}

\begin{thebibliography}{10}
\expandafter\ifx\csname url\endcsname\relax
  \def\url#1{\texttt{#1}}\fi
\expandafter\ifx\csname urlprefix\endcsname\relax\def\urlprefix{URL }\fi
\expandafter\ifx\csname href\endcsname\relax
  \def\href#1#2{#2} \def\path#1{#1}\fi

\bibitem{BERNABEI1998195}
R.~Bernabei, et~al.,
  \href{https://www.sciencedirect.com/science/article/pii/S0370269398001725}{{Searching
  for WIMPs by the annual modulation signature}}, Phys. Lett. B 424~(1) (1998)
  195.

\bibitem{Bernabei:2013xsa}
R.~Bernabei, et~al.,
  \href{https://link.springer.com/article/10.1140/epjc/s10052-013-2648-7}{{Final
  model independent result of DAMA/LIBRA-phase1}}, Eur. Phys. J. C 73 (2013)
  2648.

\bibitem{Bernabei:2018jrt}
R.~Bernabei, et~al.,
  \href{http://jnpae.kinr.kiev.ua/19.4/html/19.4.0307.html}{{First model
  independent results from DAMA/LIBRA-phase2}}, Nucl. Phys. Atom. Energy 19~(4)
  (2018) 307.

\bibitem{wrro120971}
M.~Battaglieri, et~al., \href{https://eprints.whiterose.ac.uk/120971/}{{US
  Cosmic Visions: New Ideas in Dark Matter 2017: Community Report}},
  arXiv:1707.04591 (2017).

\bibitem{PhysRevLett.118.101101}
E.~Aprile, et~al.,
  \href{https://link.aps.org/doi/10.1103/PhysRevLett.118.101101}{{Search for
  Electronic Recoil Event Rate Modulation with 4 Years of XENON100 Data}},
  Phys. Rev. Lett. 118 (2017) 101101.

\bibitem{Aprile:2018dbl}
E.~Aprile, et~al.,
  \href{https://journals.aps.org/prl/abstract/10.1103/PhysRevLett.121.111302}{{Dark
  Matter Search Results from a One Ton-Year Exposure of XENON1T}}, Phys. Rev.
  Lett. 121~(11) (2018) 111302.

\bibitem{COSINE-100:2021xqn}
G.~Adhikari, et~al.,
  \href{https://www.science.org/doi/10.1126/sciadv.abk2699}{{Strong constraints
  from COSINE-100 on the DAMA dark matter results using the same sodium iodide
  target}}, Sci. Adv. 7~(46) (2021) abk2699.

\bibitem{PhysRevD.98.062005}
D.~S. Akerib, et~al.,
  \href{https://link.aps.org/doi/10.1103/PhysRevD.98.062005}{{Search for annual
  and diurnal rate modulations in the LUX experiment}}, Phys. Rev. D 98 (2018)
  062005.

\bibitem{kwkim15}
K.~W. Kim, et~al.,
  \href{https://www.sciencedirect.com/science/article/pii/S0927650514001595}{{Tests
  on NaI(Tl) crystals for WIMP search at the Yangyang Underground Laboratory}},
  Astropart. Phys. 62 (2015) 249.

\bibitem{Adhikari:2017esn}
G.~Adhikari, et~al.,
  \href{https://link.springer.com/article/10.1140/epjc/s10052-018-5590-x}{{Initial
  Performance of the COSINE-100 Experiment}}, Eur. Phys. J. C 78~(2) (2018)
  107.

\bibitem{Antonello2021}
M.~Antonello, et~al.,
  \href{https://doi.org/10.1140/epjc/s10052-021-09098-5}{{Characterization of
  SABRE crystal NaI-33 with direct underground counting}}, Eur. Phys. J. C 81
  (2021) 299.

\bibitem{Fushimi:2021mez}
K.~Fushimi, et~al.,
  \href{https://academic.oup.com/ptep/article/2021/4/043F01/6137844}{{Development
  of highly radiopure NaI(Tl) scintillator for PICOLON dark matter search
  project}}, PTEP 2021~(4) (2021) 043F01.

\bibitem{anais2016}
J.~Amare, et~al.,
  \href{https://link.springer.com/article/10.1140/epjc/s10052-016-4279-2}{{Assessment
  of backgrounds of the ANAIS experiment for dark matter direct detection}},
  Eur. Phys. J. C 76 (2016) 429.

\bibitem{cosinus2023}
G.~Angloher, et~al., \href{https://arxiv.org/abs/2307.11139}{{Deep-underground
  dark matter search with a COSINUS detector prototype}}, arXiv:2307.11139
  (2023).

\bibitem{PhysRevD.106.052005}
G.~Adhikari, et~al.,
  \href{https://link.aps.org/doi/10.1103/PhysRevD.106.052005}{{Three-year
  annual modulation search with COSINE-100}}, Phys. Rev. D 106 (2022) 052005.

\bibitem{Amare:2021yyu}
J.~Amare, et~al.,
  \href{https://journals.aps.org/prd/abstract/10.1103/PhysRevD.103.102005}{{Annual
  Modulation Results from Three Years Exposure of ANAIS-112}}, Phys. Rev. D
  103~(10) (2021) 102005.

\bibitem{BERNABEI1996757}
R.~Bernabei, et~al.,
  \href{http://www.sciencedirect.com/science/article/pii/S0370269396800207}{{New
  limits on WIMP search with large-mass low-radioactivity NaI(Tl) set-up at
  Gran Sasso}}, Phys. Lett. B389~(4) (1996) 757.

\bibitem{Collar:2013gu}
J.~I. Collar,
  \href{https://journals.aps.org/prc/abstract/10.1103/PhysRevC.88.035806}{{Quenching
  and channeling of nuclear recoils in NaI(Tl): Implications for dark-matter
  searches}}, Phys. Rev. C 88~(3) (2013) 035806.

\bibitem{PhysRevC.92.015807}
J.~Xu, et~al.,
  \href{https://link.aps.org/doi/10.1103/PhysRevC.92.015807}{{Scintillation
  efficiency measurement of Na recoils in NaI(Tl) below the DAMA/LIBRA energy
  threshold}}, Phys. Rev. C 92 (2015) 015807.

\bibitem{Joo:2018hom}
H.~W. Joo, et~al.,
  \href{https://doi.org/10.1016/j.astropartphys.2019.01.001}{{Quenching factor
  measurement for NaI(Tl) scintillation crystal}}, Astropart. Phys. 108 (2019)
  50.

\bibitem{Bignell_2021}
L.~Bignell, et~al.,
  \href{https://dx.doi.org/10.1088/1748-0221/16/07/P07034}{{Quenching factor
  measurements of sodium nuclear recoils in NaI:Tl determined by spectrum
  fitting}}, Journal of Instrumentation 16~(07) (2021) P07034.

\bibitem{cintas2024measurement}
D.~Cintas, et~al., \href{https://arxiv.org/abs/2402.12480}{{A measurement of
  the sodium and iodine scintillation quenching factors across multiple NaI(Tl)
  detectors to identify systematics}}, arXiv:2402.12480 (2024).

\bibitem{lee2024measurements}
S.~H. Lee, et~al., \href{https://arxiv.org/abs/2402.15122}{{Measurements of low
  energy nuclear recoil quenching factors for Na and I recoils in the NaI(Tl)
  scintillator}}, arXiv:2402.15122 (2024).

\bibitem{Zurowski2022}
M.~J. Zurowski, E.~Barberio,
  \href{https://doi.org/10.1140/epjc/s10052-022-11062-w}{{Influence of NaI
  background and mass on testing the DAMA modulation}}, Eur. Phys. J. C 82
  (2022) 1122.

\bibitem{DAMA075}
R.~Bernabei, et~al.,
  \href{http://jnpae.kinr.kiev.ua/22.4/html/22.4.0329.html}{{Further results
  from DAMA/LIBRA-phase2 and perspectives}}, Nucl. Phys. Atom. Energy 22~(4)
  (2021) 329.

\bibitem{COSINE-100:2021poy}
G.~Adhikari, et~al.,
  \href{https://journals.aps.org/prd/abstract/10.1103/PhysRevD.105.042006}{{Searching
  for low-mass dark matter via Migdal effect in COSINE-100}}, Phys. Rev. D 105
  (2022) 042006.

\bibitem{neon2023}
J.~Choi, et~al.,
  \href{https://link.springer.com/article/10.1140/epjc/s10052-023-11352-x}{{Exploring
  coherent elastic neutrino-nucleus scattering using reactor electron
  antineutrinos in the NEON experiment}}, Eur. Phys. J. C 83 (2023) 226.

\bibitem{Choi:2020qcj}
J.~Choi, et~al.,
  \href{https://www.sciencedirect.com/science/article/pii/S0168900220309530}{{Improving
  the light collection using a new NaI(Tl)crystal encapsulation}}, Nucl.
  Instrum. Meth. A 981 (2020) 164556.

\bibitem{geant4}
S.~Agostinelli, et~al.,
  \href{https://www.sciencedirect.com/science/article/pii/S0168900203013688}{{Geant4—a
  simulation toolkit}}, Nucl. Instrum. Meth. A 506 (2003) 250.

\bibitem{SALDANHA201735}
R.~Saldanha, et~al.,
  \href{https://www.sciencedirect.com/science/article/pii/S016890021730311X}{{Model
  independent approach to the single photoelectron calibration of
  photomultiplier tubes}}, Nucl. Instrum. Meth. A 863 (2017) 35.

\bibitem{LEE2006201}
H.~Lee, et~al.,
  \href{https://www.sciencedirect.com/science/article/pii/S0370269305018423}{{First
  limit on WIMP cross section with low background CsI(Tl) crystal detector}},
  Phys. Lett. B 633~(2) (2006) 201.

\bibitem{doi:10.1080/14786440308521076}
J.~Moyal, \href{https://doi.org/10.1080/14786440308521076}{{XXX. Theory of
  ionization fluctuations}}, The London, Edinburgh, and Dublin Philosophical
  Magazine and Journal of Science 46~(374) (1955) 263--280.

\bibitem{DAGOSTINI1995487}
G.~D'Agostini,
  \href{https://www.sciencedirect.com/science/article/pii/016890029500274X}{{A
  multidimensional unfolding method based on Bayes' theorem}}, Nucl. Phys.
  Atom. Energy 362~(2) (1995) 487.

\bibitem{PhysRevD.107.032010}
F.~Sutanto, et~al.,
  \href{https://link.aps.org/doi/10.1103/PhysRevD.107.032010}{{Production and
  suppression of delayed light in NaI(Tl) scintillators}}, Phys. Rev. D 107
  (2023) 032010.

\bibitem{PhysRevD.93.042001}
J.~Cherwinka, et~al.,
  \href{https://link.aps.org/doi/10.1103/PhysRevD.93.042001}{{Measurement of
  muon annual modulation and muon-induced phosphorescence in NaI(Tl) crystals
  with DM-Ice17}}, Phys. Rev. D 93 (2016) 042001.

\bibitem{Adhikari_2018}
G.~Adhikari, et~al.,
  \href{https://dx.doi.org/10.1088/1748-0221/13/09/P09006}{{The COSINE-100 data
  acquisition system}}, J. Instrum. 13~(09) (2018) P09006.

\bibitem{adhikari2020lowering}
G.~Adhikari, et~al.,
  \href{https://www.sciencedirect.com/science/article/pii/S0927650521000256}{{Lowering
  the energy threshold in COSINE-100 dark matter searches}}, Astropart. Phys.
  130 (2021) 102581.

\bibitem{Tanabashi2018}
M.~Tanabashi, et~al.,
  \href{https://journals.aps.org/prd/abstract/10.1103/PhysRevD.98.030001}{{The
  Review of Particle Physics}}, Phys. Rev. D 98 (2018) 030001.

\bibitem{COSINE_SET1}
G.~Adhikari, et~al.,
  \href{https://www.nature.com/articles/s41586-018-0739-1}{{An experiment to
  search for dark-matter interactions using sodium iodide detectors}}, Nature
  564~(7734) (2018) 83.

\bibitem{hslee07}
H.~S. Lee, et~al.,
  \href{https://journals.aps.org/prl/abstract/10.1103/PhysRevLett.99.091301}{{Limits
  on Interactions between Weakly Interacting Massive Particles and Nucleons
  Obtained with CsI(Tl) Crystal Detectors}}, Phys. Rev. Lett. 99 (2007) 091301.

\bibitem{sckim12}
S.~C. Kim, et~al.,
  \href{https://journals.aps.org/prl/abstract/10.1103/PhysRevLett.108.181301}{{New
  Limits on Interactions between Weakly Interacting Massive Particles and
  Nucleons Obtained with CsI(Tl) Crystal Detectors}}, Phys. Rev. Lett. 108
  (2012) 181301.

\end{thebibliography}

\end{document}